# Band Structure of EuO/Si Spin Contact: Justification for Silicon Spintronics

Leonid L. Lev,[†, ‡] Dmitry V. Averyanov,[‡] Andrey M. Tokmachev,[‡] Federico Bisti,[†] Victor A. Rogalev,[†] Vladimir N. Strocov[+, †] and Vyacheslav G. Storchak[*,‡]

[†]Swiss Light Source, Paul Scherrer Institute, CH-5232 Villigen, Switzerland

[‡]National Research Centre "Kurchatov Institute", Kurchatov Sq. 1, 123182 Moscow, Russia

ABSTRACT. Silicon spintronics requires injection of spin-polarized carriers into Si. An emerging approach is direct electrical injection from a ferromagnetic semiconductor – EuO being the prime choice. Functionality of the EuO/Si spin contact is determined by the interface band alignment. In particular, the band offset should fall within the 0.5-2 eV range. We employ soft-X-ray ARPES to probe the electronic structure of the buried EuO/Si interface with momentum resolution and chemical specificity. The band structure reveals a conduction band offset of 1.0 eV attesting the technological potential of the EuO/Si system.

Scaling of conventional electronics faces formidable obstacles due to fundamental limits of information processing in different areas.[1] The management of power consumption and heat generation is probably the main challenge. Spintronics, among other emerging technologies, addresses this issue by employing spin degrees of freedom. Metallic spintronics devices provide efficient data storage[2] but transistor action requires semiconductors.[3] In particular, Si spintronics[4,5] is especially appealing due to the dominant role of Si in the modern electronics. As Si is non-magnetic, it requires creation of spin-polarized carriers in the system. Direct spin injection from a ferromagnetic (FM) metal into Si is ineffective due to the well-known impedance mismatch problem.[6] Insulating tunnel barriers between an FM metal and silicon promote spin injection into bulk Si[7,8] and Si nanotubes.[9] Alternatives based on hot electrons,[10] thermal[11] and acoustic[12] spin injection, spin pumping[13] are also available. However, spin injection characteristics sufficient for technological applications are yet to be demonstrated. Perhaps the solution is in the choice of material integrated with Si and the quality of the integration itself: even FM tunnel contacts can be dramatically improved by replacing the material of the barrier[14] or advances in the growth procedure.[15]

The impedance mismatch problem can be alleviated by using half-metallic injectors.[16] However, the most straightforward approach is based on electrical injection of spins into Si from an FM semiconductor.[17] Europium monoxide EuO is considered the best candidate for such a heterostructure due to the unique combination of physical and technological properties. First, it is compatible with silicon: EuO is a binary compound with the rock-salt structure thermodynamically stable in contact with silicon;[18] band gap of EuO (1.1 eV) matches that of Si. Secondly, remarkable bulk properties of EuO – colossal magnetoresistivity of about 6 orders of magnitude in 2 T, metal-insulator transition accompanied by up to 15 orders of magnitude change in resistivity, very strong magneto-optic effects, *etc.* – open ways for multi-functional devices. Thirdly, magnetic properties of EuO can be tuned by doping,[19,20] strain[21] or optical pumping.[22] Finally, EuO is a magnetically homogeneous on the nanoscale[23] source of almost 100% spin-polarized electrons (due to enormous exchange splitting $2\Delta E_{ex} \sim 0.6$ eV)[24] inviting spin-filter applications.[25]

These remarkable properties prompt numerous attempts to integrate EuO with different substrates[26-29] as well as computational studies of the electronic structure of EuO[20,30] and magnetic effects coming from interfacing EuO with other materials.[31-33] However, direct epitaxial integration of EuO with Si turns out to be a notoriously difficult technological problem[34-36] not only because of the significant lattice mismatch of 5.6%. Integration of crystalline oxides with Si is always challenging[37] but EuO synthesis is aggravated by interfacial chemistry: formation of higher oxides $Eu_3O_4$ and $Eu_2O_3$ as well as reactions of both Eu and oxygen with the substrate. As a result, a layer of side products separates EuO and Si[35,36] preventing formation of a direct spin contact necessary for spintronics applications.

Recently, we proposed a new template for the growth of EuO (and other oxides) on Si suppressing unwanted chemistry at the interface and leading to direct epitaxial integration of the semiconductors.[38,39] Moreover, analytical electron microscopy reveals that the interface engineering ensures an atomically sharp EuO/Si interface.[40] Due to these advances in the growth methodology a clean direct EuO/Si interface free of any intermediate layer becomes available for studies of its functionality for spintronics applications.

The band structure at the interface (band bending and the resulting band offset) is the most important characteristic for the intended carrier injection. Angle-resolved photoemission spectroscopy (ARPES) is a direct technique to study the electronic structure of the surface layer with resolution in electron momentum **k**. ARPES has been previously used to probe the electronic structure of pristine EuO[41] and Gd-doped EuO.[42,43] The measurements reveal the Eu 4$f$ band and the origin of the magnetism of EuO,[41] the appearance of new occupied states in the spectrum near the Fermi level $E_F$ due to Gd doping,[42] as well as evolution of the electronic structure through the FM metal-insulator transition.[43] However, the electronic structure of the EuO/substrate interface – a major ingredient of a spintronic device – has not been addressed so far due to extremely small probing depth of the conventional ARPES with photon energies $hv$ around 100 eV employed in Refs. 41-43. The EuO/Si interface is deeply buried within the grown structure due to finite EuO film thickness and additional thickness of unavoidable capping layer protecting EuO from the atmosphere. Thus, the problem is highly challenging for ARPES.

Here we present the first study of the buried EuO/Si interface with soft-X-ray (SX) ARPES techniques employing synchrotron radiation with $hv$ around 1 keV. The experiment hinges on a successful synthesis of ultrathin (5 monolayers) EuO films epitaxially integrated with Si and protected by a minimal (15-20 Å) capping layer of $SiO_x$. Remarkably, the use of soft X-rays clearly exposes the bulk band dispersion of the substrate through about 30 Å of the combined EuO and $SiO_x$ overlayer. In combination with the energy position of the $Eu^{2+}$ levels determined with resonant photoexcitation, it allows for a detailed study of the EuO and Si bands at the interface. The experimental band alignment is highly encouraging for Si spintronics, suggesting that efficient injection of spin-polarized carriers from EuO into Si is viable.

The choice of ARPES techniques capable of assessing the buried interface is of paramount importance. Traditional ARPES employs the vacuum-ultraviolet (VUV) region making it an extremely surface sensitive probe: the probing depth characterized by the electron inelastic mean free path λ is only several Å.[44] To avoid the measurement of surface-associated artifacts *in situ* sample transfer between the growth and analysis chambers is preferable. However, the vacuum coupling is technically difficult and the procedure precludes macroscopic characterization of the film before the measurement. *Ex situ* VUV-ARPES study of the EuO/Si system would require surface cleaning such as ion sputtering inevitably destroying the interface.

The only reliable way to enhance bulk and buried interface sensitivity is to utilize higher photon energies, where λ grows as [photoelectron kinetic energy]$^{3/4}$. Higher energies bring further benefits: the uncertainty principle suggests that smearing in surface-perpendicular projection of **k**[45] would decrease as 1/λ; the high kinetic energy of photoelectrons ensures that the final state can be approximated as a free electron.[46] Although successful hard X-ray (HX-) ARPES studies in the multi-keV range (boosting λ up to 100 Å) have been reported for a number of materials[46,47] the method is probably not advisable for the study of the EuO/Si system. First, the energy resolution gradually deteriorates with the increase of the photon energies. In the limit of high energy the creation and annihilation of phonons can smear out the **k** specification of the final state.[48] Moreover, in the multi-keV range the photoelectron momentum is not negligible and recoil effects become significant, especially for lighter atoms. Finally, the applications of HX-ARPES are much impeded by progressive reduction of the VB cross-section.[49] All this taken into account, SX-ARPES in the *hν* range around 1 keV seems to be the golden mean between VUV- and HX-ARPES for studies of the buried EuO/Si interface and similar systems. Moreover, monitoring of resonant enhancement of valence states at the *L*- and *M*-absorption edges in this energy range enables determination of the element-specific band structure. However, the problem requires pushing SX-ARPES to the limits of its energy resolution and probing depth. As for the sample prerequisites, the limitations of SX-ARPES set up severe restrictions on the film thickness. Thus, a fine balance between requirements stemming from the growth and ARPES methodologies should be found.

The growth of ultrathin films of EuO is exceptionally difficult: standard approaches lead to layers of alien side product phases with thicknesses exceeding the required thickness of the EuO layer.[35,36] Moreover, annealing procedures may lead to intermixing at the interface. Therefore, we employ a meticulously tailored growth scheme based on principles developed in Refs. 38-40. The films are grown in Riber Compact 12 system for molecular beam epitaxy of oxides. The substrates are *n*-Si (001) wafers with miscut angles not exceeding 0.5°. Doped *n*-Si is used to eliminate charging effects in the subsequent photoemission (PE) experiments. The natural surface oxide is removed by heating to reveal the standard 2×1 reconstruction of the clean Si (001) surface, monitored *in situ* with reflection high-energy electron diffraction (RHEED). This surface is highly reactive and requires protection before the oxide growth.

The standard approach is to develop a submonolayer surface silicide.[50] In practice, saturated surface strontium silicide $SrSi_2$ with 1×2 reconstruction is used (in the case of the EuO growth isomorphous surface silicide $EuSi_2$ is more appropriate). Regretfully, such protection is not sufficient for the growth of EuO directly on Si.[36,38] Therefore, we employ another Eu surface silicide phase with a higher Eu content, seen as 1×5 reconstruction on RHEED images.[38] It is formed when the Si surface is exposed to a flux of Eu (4N) at 660 °C. One should take into account that the type and structure of the surface silicide used for protection may affect the interface band alignment[51] although structural studies suggest oxidation of M-Si bonds and incorporation of the interfacial metal layer into the oxide system.[52] EuO films are grown on the protected surface at a temperature of 340±10 °C, an oxygen (6N) pressure of $6·10^{-9}$ Torr and a temperature of 500±10 °C of the Eu effusion cell. The thickness of the films is limited to 5 monolayers – the minimum required for the lattice mismatch relaxation in the film.[38,50] The FM transition temperature of these extremely ultrathin films is shifted to 10 K (measured by SQUID) due to enormous relaxation of EuO lattice constant 5.14 Å to 5.43 Å of the Si substrate. The film needs a capping layer because EuO reacts with atmospheric $O_2$ and $H_2O$. The routinely used protection by amorphous $Eu_2O_3$[38,39] is not suitable because it prevents control of the thickness of the EuO film and a large amount of $Eu^{3+}$ hinders interpretation of ARPES spectra. Therefore, we employ a capping layer of amorphous $SiO_x$ with a thickness of about 20 Å. Our experience based on X-ray diffraction and transmission electron microscopy studies is that $SiO_x$ layer of such thickness is sufficient for protection of the most part of the film but due to possible unevenness of the capping layer some regions of EuO top surface may become oxidized to $Eu_2O_3$. Fortunately, this is not an obstacle for SX-ARPES study of the band structure at the buried EuO/Si interface.

The general scheme of our SX-ARPES experiment on the $SiO_x$/EuO/Si structures is presented in Fig. 1*a*. Monochromatic X-rays produced by the synchrotron eject photoelectrons from the $SiO_x$, EuO and Si layers. The photoelectron analyzer detects their intensity distribution $I_{PE}(E_k,\theta)$ as a function of photoelectron kinetic energy $E_k$ and emission angle $\theta$ which render into binding energy $E_b$ and momentum **k** of these electrons back in the valence band (VB) of the sample to yield its band structure $E(\mathbf{k})$ as the electron binding energy $E_b$ depending on momentum **k**.

Our experiments have been carried out at the SX-ARPES endstation[53] of the ADRESS beamline[54] at the Swiss Light Source (Paul Scherrer Institute, Switzerland). We used *p*- polarization of incident X-rays. The experimental geometry[53] with the slit of the photoelectron analyzer PHOIBOS-150 oriented along the incident beam sets the **k**-axes orientation relative to the Brillouin zone (BZ) of Si and EuO having fcc crystal structure as shown in Fig. 1b: the projection $k_x$ is directly measured through the emission angle along the analyzer slit, $k_y$ is varied by tilt rotation of the sample, and $k_z$ through $hv$. The experiments are carried out at the lowest available temperature of 12 K to quench thermal effects detrimental to the coherent **k**-resolved spectral component at high photoelectron energies.[48] On the other hand, this temperature is above $T_C$ of our ultrathin film samples. The combined (beamline and analyzer) energy resolution was ~0.2÷0.25 eV. The position of $E_F$ was monitored by measurements at Au foil in electrical contact with the sample. No photon flux dependent charging effects were detected due to small thickness of the film and *n*-doping of the Si substrate. The resonant SX-ARPES studies were complemented by X-ray adsorption spectra (XAS) measurements carried out in the total electron yield. First, it is necessary to establish that the SX-ARPES probing depth is sufficient to penetrate through the film and reach the EuO/Si interface and Si substrate. The conclusion can be drawn from the PE response of the Si 2p core levels. Fig. 2 shows the angle-integrated PE spectrum measured at $hv$ = 1300 eV. Being sensitive to the chemical state of Si atoms, this spectrum shows the characteristic $2p_{3/2}$-$2p_{1/2}$ spin-orbit split doublet at $E_b$ = -98.7 and -99.5 eV identifying neutral Si atoms in the Si substrate, and a broad hump around $E_b$ = -103.5 eV identifying positively charged Si ions in different oxygen environments in the non-stoichiometric $SiO_x$ capping layer (see, for example, Ref. 55). The observation of a signal from $Si^0$ is remarkable evidence that our SX-ARPES probing depth is sufficient to penetrate through the $SiO_x$ capping and reach the EuO/Si interface and Si substrate. Moreover, quantitative analysis of the relative integral intensities of the peaks (see Supporting Information) has shown that within a practical sensitivity limit of 2% the probing depth extends as much as about 60 Å into the Si substrate.

The next step is to get information on the EuO spin injector layer. First, we used XAS with its large probing depth and elemental and chemical state specificity, allowing for discrimination of Eu ions with different oxidation number. XAS spectrum at the Eu $3d_{5/2}$ adsorption edge (Fig. 3a) reveals a dominant peak identified as $Eu^{2+}$.[56] It determines the prevailing stoichiometry of the EuO film, where most of EuO is intact and not oxidized by air. The $Eu^{2+}$ peak is accompanied by smaller ones which can be identified as $Eu^{3+}$.[56] The small admixture of $Eu^{3+}$ is not surprising since the amorphous capping layer is intentionally made extremely thin to enable the SX-ARPES study, with the side effect being the presence of certain fraction of spatial regions with insufficient protection of EuO.

Furthermore, we probed the VB region of the EuO layer with the element and chemical state specific resonant PE[57,58] to get detailed information on the Eu contributions in different oxidation states. Fig. 3b shows the resonant PE map of angle-integrated $I_{PE}(E_b,hv)$ measured through the VB region under variation of $hv$ across the Eu $3d_{5/2}$ absorption edge. Tuning $hv$ to different peaks of the XAS spectrum resonantly enhances the PE signal from Eu ions with the corresponding oxidation state, allowing thus chemical state resolution of the VB. In particular, the XAS peak at ~1128.5 eV, corresponding to $Eu^{2+}$ ions, manifests itself as the strong $I_{PE}(E_b,hv)$ resonant peak at $E_b$ around -2.7 eV. The weaker XAS peaks, corresponding to the $Eu^{3+}$ ions discussed above, manifest themselves as the weaker peaks in the $E_b$ region from -12 to -6 eV. Figs. 3c – 3e show the angle-resolved PE images $I_{PE}(E_b,k_x)$ measured at the $Eu^{2+}$ and two strongest $Eu^{3+}$ resonances. The narrow band centered around -2.7 eV is the $Eu^{2+}$ derived $^7F_0 - {^7F_6}$ multiplet. We do not detect any notable dispersion of $Eu^{2+}$ or $Eu^{3+}$ states, in accord with the highly localized character of Eu 4*f* electrons. Possible extremely weak dispersion effects near the top of the $Eu^{2+}$ multiplet, caused by admixture of delocalized O 2*p* orbitals and reported in the previous VUV-ARPES study,[41] would be completely suppressed in our data because of vanishing photoexcitation cross-section of O 2*p* compared to Eu 4*f*, especially at the Eu 3*d* resonance. The narrow energy width of the $Eu^{2+}$ peak (Fig. 3c) certifies that charging effects in the thin EuO film are negligible.

In our *hv* range around 1100 eV, photoelectrons from the VB region have $E_k$ sufficiently high to escape into vacuum through the $SiO_x$ and EuO layers and bring spectroscopic information about the Si substrate. With the bulk lattice parameters of Si, tuning *hv* to 1120 eV brings $k_z$ (corrected for the X-ray photon momentum[53]) to the Γ-point of its 3D bulk BZ. Fig. 4a shows the resulting raw PE image $I_{PE}(E_b,k_x)$ recorded at $k_y = 0$, which corresponds therefore to the ΓKX direction. The image is dominated by $Eu^{2+}$ and $Eu^{3+}$ structures, and intense non-dispersive background coming mostly from the amorphous $SiO_x$ layer and photoelectrons from the Si substrate which smear their **k**-definition during quasielastic scattering in the amorphous $SiO_x$ layer during their escape to vacuum. We can however discern sharp dispersions on top of the background. This spectral component can be enhanced by subtracting the non-dispersive one, shown in Fig. 4d, which is determined by integration of the $I_{PE}(E_b,k_x)$ image over the intercepted $k_x$ interval. The enhanced $E(\mathbf{k})$ in Fig. 4g immediately identifies the textbook light-hole and heavy-hole bands of bulk Si along ΓKX informed by photoelectrons penetrating through the $SiO_x$ and EuO layers. Detuning *hv* from 1120 eV results in a downward dispersion of the observed bands (not shown here for brevity) confirming the Γ-point location of our $k_z$. Furthermore, Figs. 4b, 4e, 4h and 4c, 4f, 4i show similar results acquired at the same *hv* = 1120 eV (and thus $k_z$ in the Γ-point) but with different sample tilt angles (and thus different $k_y$). They reflect the evolution of $E(\mathbf{k})$ along off-symmetry directions parallel to ΓKX. We note that the sharpness of the observed Si bands rules out any significant space charge effects in our *n*-doped Si substrate. The response of the Si substrate can also be illustrated by maps of $I_{PE}(k_x,k_y)$ measured under continuous variation of the sample tilt (see Supporting Information).

Concluding our analysis of SX-ARPES experimental data, we note that the observation of the coherent band structure signal from the Si substrate buried under a 30 Å overlayer of EuO and $SiO_x$ is a remarkable example of the probing capability of this technique.[59] The standard X-ray photoelectron spectroscopy is hardly suitable in our case of the complex $SiO_x$/EuO/Si film because, as apparent from Figs. 4b, 4e, 4f, the angle integration completely erodes the dispersive Si signal. Most important, however, fixed *hv* of the laboratory X-ray sources permit neither resonant PE to determine the VB chemical composition, nor tuning $k_z$ required for navigation in 3D **k**-space of the Si substrate.

Analysis of our results in Fig. 4g allows us to determine the main quantitative result of this work, the band offset at the EuO/Si interface critical for the spin injecting functionality of the EuO/Si spin contact. Fig. 5a shows the energy distribution curve (EDC) extracted from the image in Fig. 4g along $k_x = 0$ and corresponding to the Γ-point of the bulk BZ. The two peaks in this EDC correspond to the VB maximum (VBM) of EuO and that of Si. Their energy difference estimates the band offset $\Delta E_V$ between the two VBMs. However, in our case its estimate on the EuO side is aggravated by the multiplet structure of the $Eu^{2+}$ derived 4*f* band forming the VBM. To determine the energy position of the upper $^7F_0$ level, we fitted the experimental $Eu^{2+}$ peak with a sequence of (experimental resolution limited) Gaussians, describing the $^7F_0 - ^7F_6$ levels with their weights equal to 2J+1 (J=0..6) and the width and energy separation treated as fitting parameters. This fit placed the $^7F_0$ multiplet level at -2.0 eV. The difference between this level and the VBM of Si positioned at -1.2 eV yields $\Delta E_V \sim 0.8$ eV, as indicated at the band diagram in Fig. 5b. We emphasize that this exact result can only be achieved by combining the resonant and **k**-selective abilities of SX-ARPES, both requiring synchrotron radiation sources with tunable *hv*.

The injection of spins at the EuO/Si interface is expected to proceed from the conduction band (CB) of electron-doped EuO into that of Si. Although the corresponding CB offset $\Delta E_V$ is not directly accessed by ARPES, in our case of a contact between paramagnetic EuO and Si it is unlikely to differ significantly from our $\Delta E_V \sim 0.8$ eV because the experimentally established optical bandgaps of EuO in paramagnetic state and Si are virtually the same (1.1 eV).[60] Furthermore, electron doping of EuO either by oxygen deficiency or by trivalent rare-earth ions on the level of a few percent does not much affect the bandgap. In practice, the spin injection requires ferromagnetic EuO where the exchange splitting of the CB minimum reduces the band gap to ~0.9 eV[60] and therefore increases the $\Delta E_C$ to ~1.0 eV, as shown in Fig. 5b.

The determined $\Delta E_C$ value makes the EuO/Si interfaces particularly suitable for modern electronics. The characteristic range of optimal supply voltage makes a rather narrow window: transistors cannot operate efficiently, on the one hand, at low supply voltage due to circuit reliability and noise issues and, on the other hand, at high voltage because integrated circuit power consumption depends quadratically on the supply voltage. Therefore, the supply voltage in electronic devices has been steadily decreasing for many years, but has recently stabilized at 0.5 – 2 V.[1] Quite remarkably, the determined band offset of the EuO/Si spin contact structure falls within this optimal range which justifies its application in silicon spintronics.

Summarizing, we have explored the band structure of buried EuO/Si spin contacts. Our study used a novel synchrotron radiation based technique of SX-ARPES with *hv* around 1 keV, which offers a unique combination of an enhanced probing depth, selectivity in 3D electron momentum space, and elemental and chemical state specificity achieved through resonant photoexcitation. The EuO/Si contacts, tailored to match the probing depth of SX-ARPES, were grown with a 13 Å thick EuO layer and capped with a 17 Å thick $SiO_x$ layer. The electronic structure characteristic most critical for the spin injection functionality of the EuO/Si interface, the $\Delta E_C$ band offset, is evaluated as 1.0 eV. Remarkably, this value falls within the optimal supply voltage range of modern electronic devices. This result justifies therefore application of EuO as spin injector into silicon.

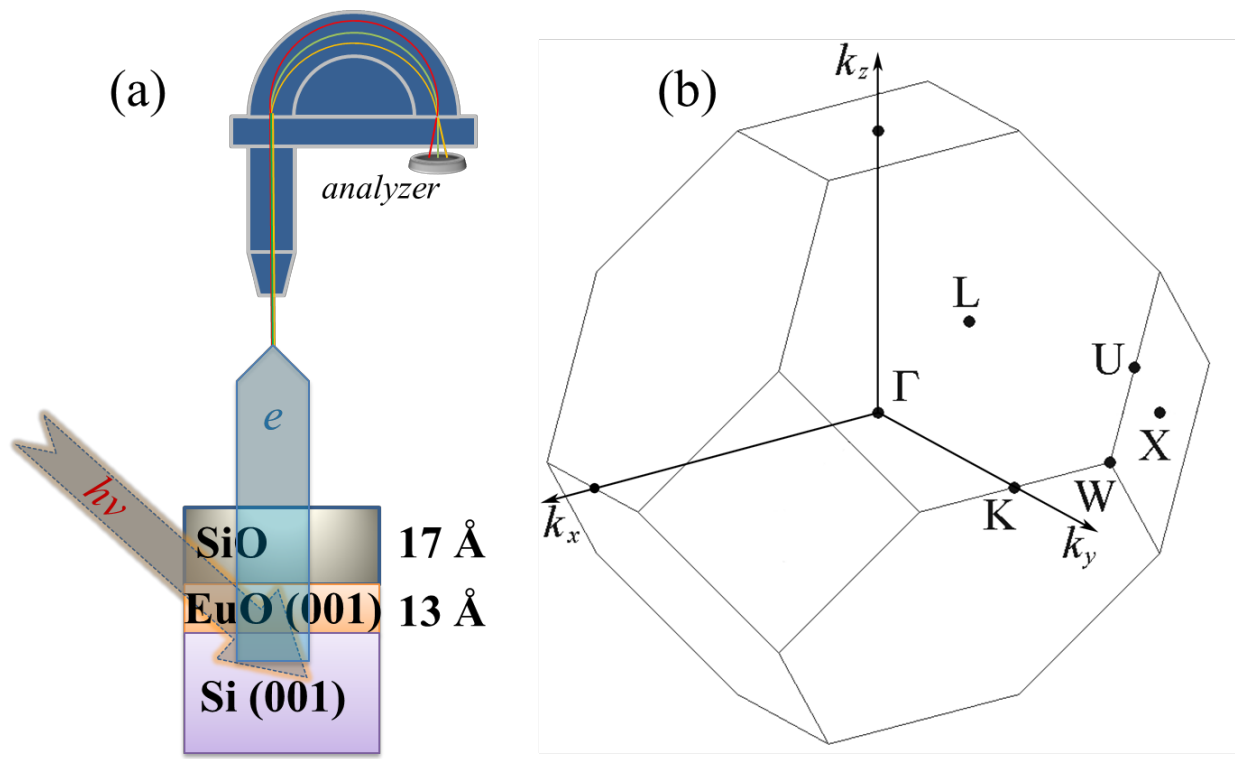

**Figure 1.** a) A sketch of SX-ARPES study of the EuO/Si interface. b) Orientation of the experimental setup **k**-axes with respect to the Brillouin zone of Si.

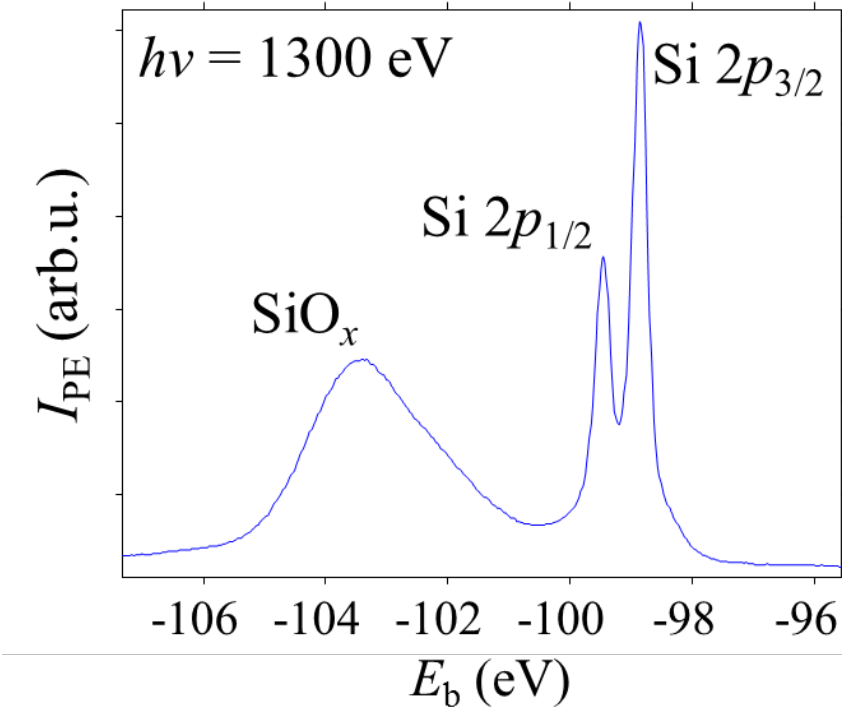

**Figure 2.** Si 2p core level spectrum measured at $h\nu = 1300$ eV. The doublet at higher $E_b$ comes from neutral Si atoms of the substrate, while the broad peak at smaller $E_b$ originatesfrom positively charged Si ions of the capping layer.

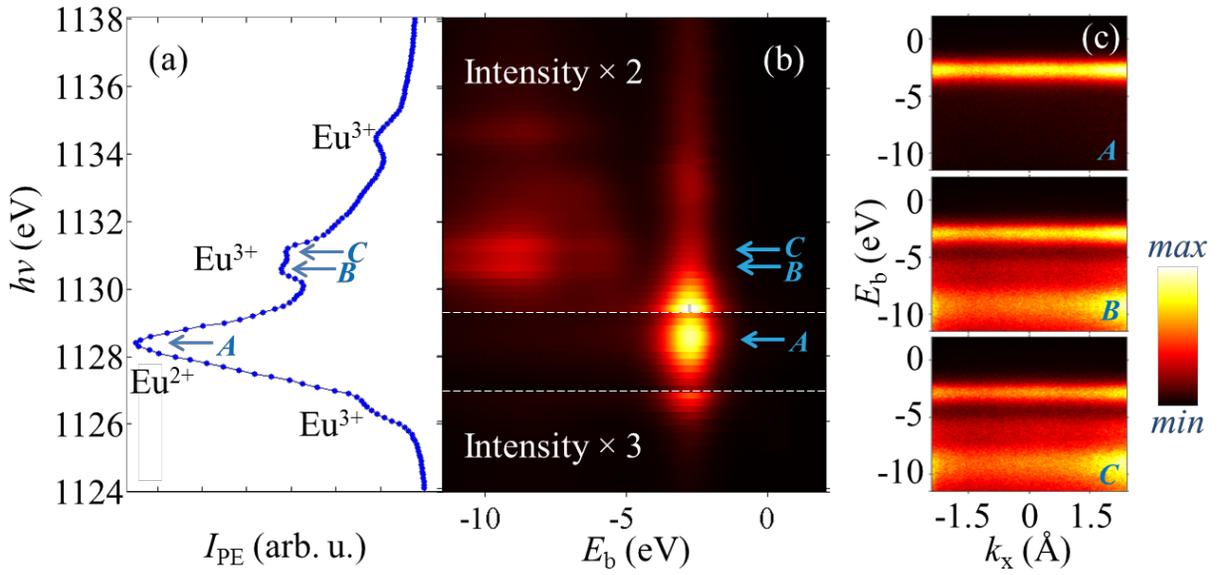

**Figure 3.** Resonant SX-ARPES at the Eu $3d_{5/2}$ edge. a) XAS spectrum showing peaks corresponding to different oxidation states of Eu ions. b) Angle-integrated $I_{PE}$ map showing the resonating Eu valence states. The map is split into three parts with different amplification of $I_{PE}$. c) Angle-resolved PE images at the main XAS peaks corresponding to $h\nu$ values A) 1128.4 eV, B) 1130.6 eV and C) 1131.0 eV.

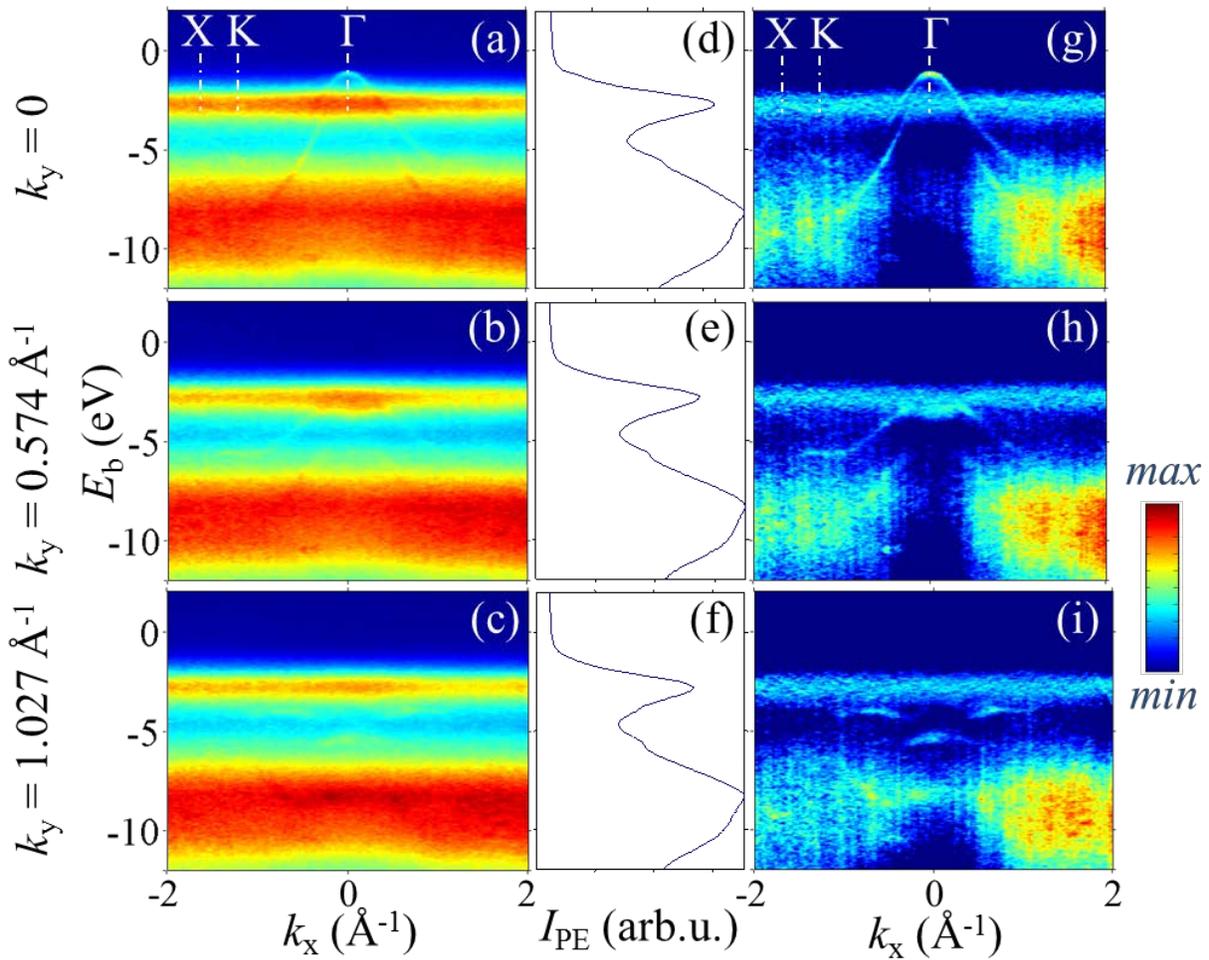

**Figure 4.** Bands of bulk Si as spectroscopic fingerprints of the Si substrate. a)–c) PE images recorded at $hv = 1120$ eV for different sample tilts corresponding to $k_y = 0$ (ΓKX direction), $k_y = 0.574$ Å-1 and $k_y = 1.027$ Å-1, respectively. d)–f) Angle-integrated spectra a)–c), respectively. g)–i) PE images a)–c) after corrections suppressing the non-dispersive contribution, respectively.

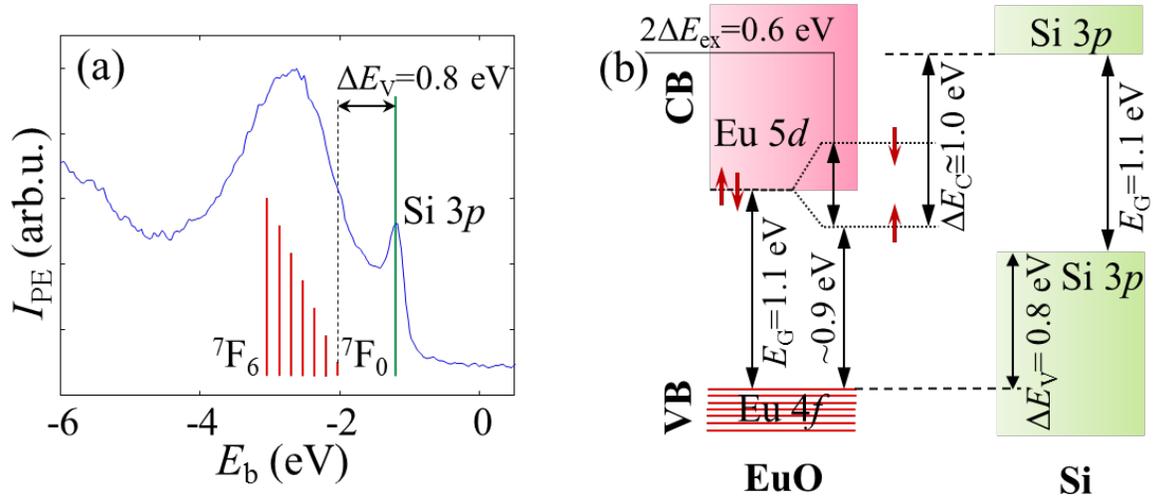

**Figure 5.** a) Experimental EDC extracted from Fig. 4a ($h\nu = 1120$ eV) corresponding to the Γ-point in Si. The two peaks are the VBM of EuO ($^7F_0 - {}^7F_6$ multiplet) and that of Si. Indicated is their offset $\Delta E_V = 0.8$ eV. b) Band diagram at the EuO/Si interface. It shows that the conduction band offset $\Delta E_C$ between EuO in the FM state and Si is 1.0 eV.

## ASSOCIATED CONTENT

**Supporting Information**.

Estimates of Si probing depth in the sample of EuO on Si capped with $SiO_x$. Maps of $I_{PE}(k_x,k_y)$ at different $E_b$ and a video showing their continuous evolution with $E_b$. This material is available free of charge via the Internet at http://pubs.acs.org.

## AUTHOR INFORMATION

**Corresponding Authors**

*E-mail: mussr@triumf.ca (materials physics)

+E-mail: vladimir.strocov@psi.ch (spectroscopy).

**Author Contributions**

The manuscript was written through contributions of all authors. All authors have given approval to the final version of the manuscript.

**Notes**

The authors declare no competing financial interest.

## ACKNOWLEDGMENT

This work is partially supported by NRC "Kurchatov Institute", Russian Foundation for Basic Research through grant 16-07-00204, Russian Science Foundation through grant 14-19-00662, and Swiss National Science Foundation through project 200021_146890. We thank D. V. Vyalikh and M. B. Tsetlin for discussions, T. Schmitt for the beamline support, and A. L. Vasiliev and A. N. Taldenkov for sample characterizations.

For Table of Contents Use Only

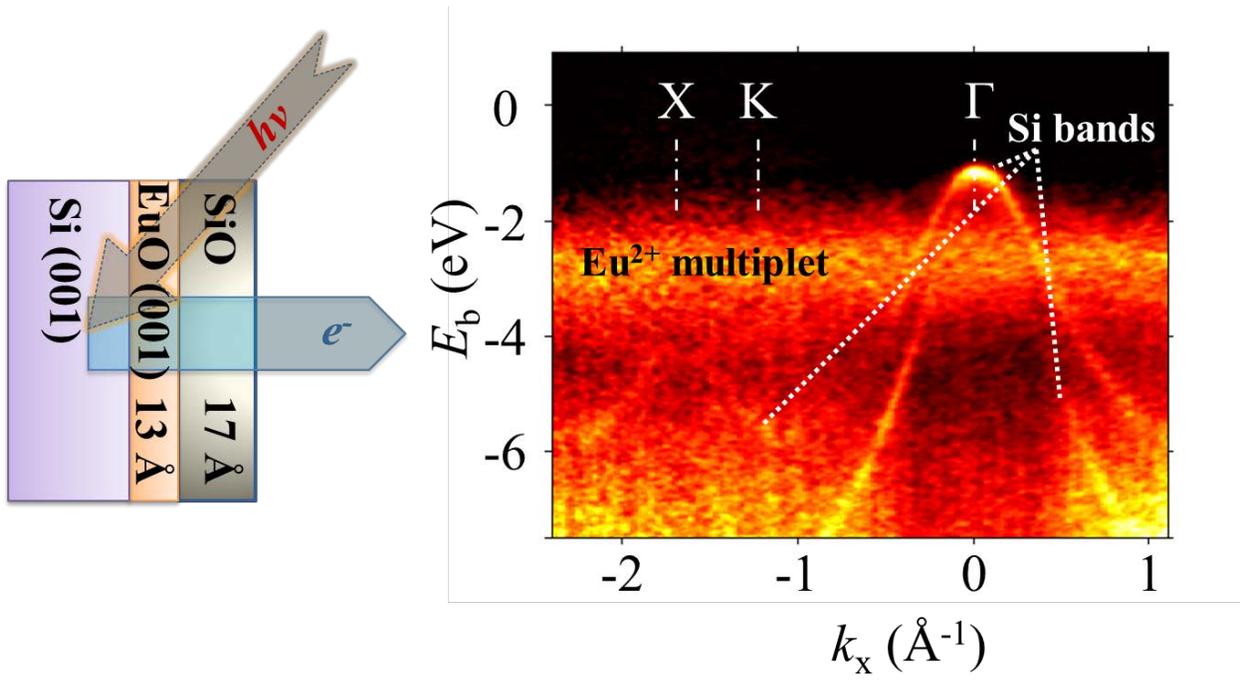